# Terahertz-Driven Phonon Upconversion in SrTiO$_3$

M. Kozina,[1] M. Fechner,[2] P. Marsik,[3] T. van Driel,[1] J.M. Glownia,[1] C. Bernhard,[3] M. Radovic,[4] D. Zhu,[1] S. Bonetti,[5] U. Staub,[4] and M.C. Hoffmann[1]

[1]*Linac Coherent Light Source, SLAC National Accelerator Laboratory, Menlo Park, California 94025, USA*

[2]*Condensed Matter Department, Max Planck Institute for the Structure and Dynamics of Matter, CFEL, 22761 Hamburg, Germany*

[3]*Department of Physics, University of Fribourg, Chemin du Musée 3, CH-1700 Fribourg, Switzerland*

[4]*Swiss Light Source, Paul Scherrer Institut, 5232 Villigen PSI, Switzerland*

[5]*Department of Physics, Stockholm University, SE-106 91 Stockholm, Sweden*

**Abstract**

Direct manipulation of the atomic lattice using intense long-wavelength laser pulses has become a viable approach to create new states of matter in complex materials. Conventionally, a high frequency vibrational mode is driven resonantly by a mid-infrared laser pulse and the lattice structure is modified through indirect coupling of this infrared-active phonon to other, lower frequency lattice modulations. Here, we drive the *lowest* frequency optical phonon in the prototypical transition metal oxide SrTiO$_3$ well into the anharmonic regime with an intense terahertz field. We show that it is possible to transfer energy to *higher* frequency phonon modes through nonlinear coupling. Our observations are carried out by directly mapping the lattice response to the coherent drive field with femtosecond x-ray pulses, enabling direct visualization of the atomic displacements.

In many complex condensed matter systems, small changes to the crystal lattice structure can drastically alter electronic properties including conductivity, polarization, orbital, charge, and spin-order[1–6]. Exploration of a phase diagram conventionally requires changing external parameters such as chemical composition, temperature, static pressure, strain or magnetic fields[6–9]. Additionally, optical excitation using ultrashort laser pulses has been used to dynamically change the electronic and crystal structure[2–5,10,11]. This method allows access to non-equilibrium states of matter that show unconventional and unique properties. Recent developments in mid-infrared laser sources have enabled a new route to control material properties: resonant excitation of phonon modes to dynamically alter the lattice structure and phonon-phonon coupling are exploited to coherently control inaccessible (i.e. non-infrared-active) phonon modes[12]. This approach is often dubbed *"nonlinear phononics"* and requires precisely tuned mid-infrared radiation and impulsive excitation[13].

Here we show a completely novel approach to alter the lattice structure by driving the lowest frequency optical phonon of the incipient ferroelectric $SrTiO_3$ (STO) into the strongly nonlinear regime by using intense broadband terahertz (THz) radiation. Specifically, we demonstrate that it is possible to transfer energy from a *low-frequency* phonon mode to *higher-frequency* modes. We anticipate that our approach will open up a deeper understanding of quantum materials through direct manipulation of the lattice structure and the utilization of dynamic coupling of vibrational modes.

STO has a low-frequency zone-center transverse optical (TO) phonon that is infrared (IR) active and highly temperature dependent[14–16]. This so called *soft-mode* phonon mediates the ferroelectric phase transition in primarily displacive ferroelectrics such as $BaTiO_3$ or $PbTiO_3$[17–19].

In bulk STO, while the soft mode shifts to lower frequencies with decreasing temperature, the system never reaches a ferroelectric state due to quantum fluctuations[20]. The temperature dependence and large anharmonicity make this mode ideally suited for probing nonlinear phonon interactions. While all-optical techniques can yield indirect evidence of excited phonons[21–23], x-ray diffraction measurements of the structural evolution of the STO unit cell provide quantitative information about the atomic motion. Recent developments in hard x-ray free electron lasers enable these ultrafast structural determinations[24–26].

We illustrate the principle of our experiment in Fig. 1. A strong-field THz pulse containing spectral components from approximately 0.2 to 2.5 THz interacts with the soft mode phonon of an STO thin film (50 nm). The frequency of the soft mode phonon can be tuned in the range of 1.5 to 2.5 THz by varying the sample temperature[27]. The relative positions of the atoms in the unit cell (Sr, turquoise; Ti, red; O, white) is monitored by diffraction of femtosecond x-ray pulses with duration (~30 fs) considerably shorter than a cycle of the THz field (see Methods below).

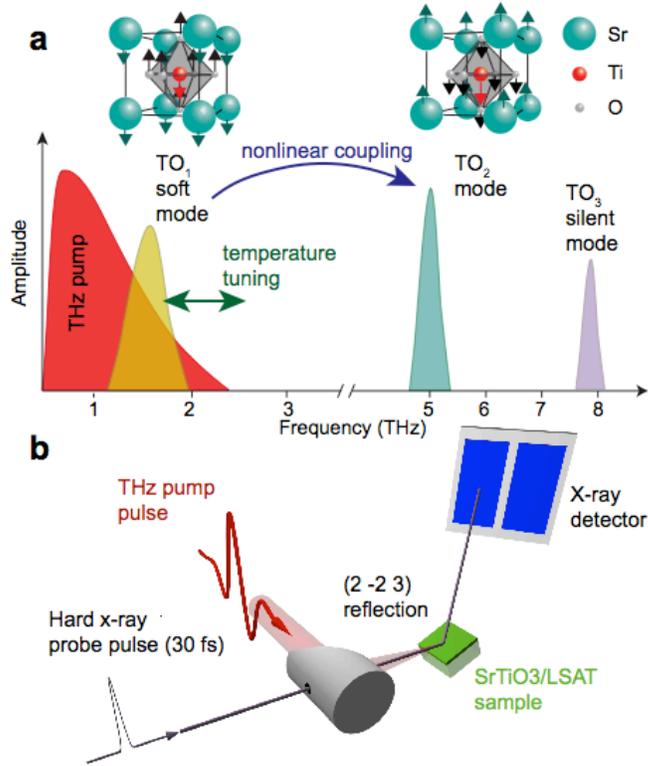

**Figure 1: Phonon frequency spectrum and experimental overview.** **a,** Strong THz radiation (red) interacts with the STO soft mode phonon (yellow). The degree of resonant overlap can be tuned by temperature. Energy is exchanged with higher frequency phonon modes (turquoise, purple) through nonlinear coupling. The STO unit cell and two lowest-frequency zone-center TO eigenmodes are indicated at the top of the figure. **b,** Phonon motion is probed in the time domain with ultrafast x-ray diffraction in reflection geometry.

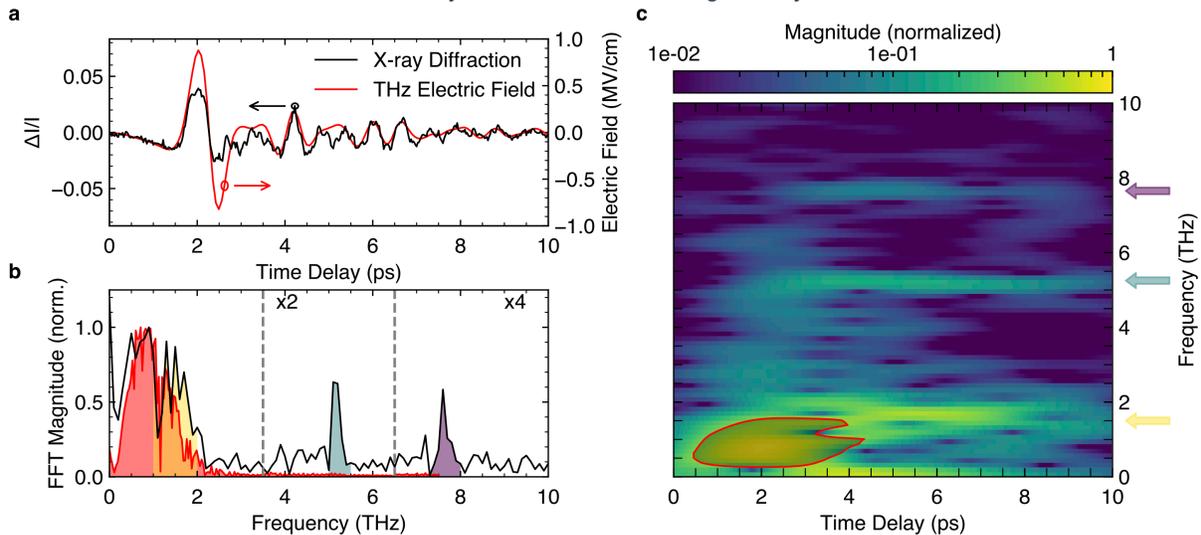

**Figure 2: Time-resolved x-ray diffraction data at 100 K.** **a,** Intensity change of the (2 -2 3) peak in STO (black) as function of time delay compared to the THz pump field (red). The traces are shifted to overlap in time as their relative timing is only known within 1 ps. **b,** Fourier transform of the time domain x-ray data shows distinct peaks, identified as the soft mode at 1.5 THz, the $TO_2/LO_1$ mode at 5.15 THz and the $TO_3/LO_2$ mode at 7.6 THz. The spectrum of the THz excitation pulse is shown in red. **c,** A short-time Fourier transformation with a 2.5 ps Kaiser window reveals the immediate onset of oscillations. The red shaded area indicates the spectrogram of the THz pump pulse (contour at 30% of peak magnitude).

Figure 2a shows the time-resolved change in the x-ray scattering (trXRD) intensity of the (2 -2 3) STO peak at ~100 K (black), overlaid with the electric field of the incident THz radiation ($E_{THz}$, red). The trXRD and $E_{THz}$ signals are shown with their peaks overlapped in time since their relative time of arrival is only known with a precision of about 1 ps. At early times (<1.5 ps) the trXRD response appears to closely follow $E_{THz}$ but then shows strong saturation and a number of additional high frequency features after the main peak which do not appear in $E_{THz}$. The spectral content of both data sets is shown in Fig. 2b normalized to a peak value of unity. At low frequencies (< 1 THz), the spectral content of $E_{THz}$ and trXRD are very similar, suggestive that the motion of the ions in the STO film primarily follows the driving THz field [28]. However, in the frequency band between one and two THz (shaded in yellow), the trXRD signal has a much larger magnitude than $E_{THz}$. Moreover, while the driving field spectral content is essentially zero at frequencies higher than 3 THz, there are clear features in the trXRD signal at 5.15 THz and 7.6 THz (shaded turquoise and purple, respectively). We interpret the spectral weight in the 1-2 THz region as a signature of the resonant excitation of the soft mode, which is well known to have strong IR absorption at these frequencies below room temperature and is strongly damped[29,30]. The soft mode can be excited directly, since there is still significant spectral overlap between the THz pump pulse and the absorption spectrum of the phonon[31]. We identify the features at 5.15 THz and 7.6 THz as two pairs of zone-center transverse and longitudinal optical (TO, LO) phonons. The frequencies of the $TO_2$ (5.3 THz) and $LO_1$ (5.2 THz) modes are very close in STO[32,33] and so we cannot distinguish them in our data, whereas the $TO_3/LO_2$ modes (7.8 THz) are degenerate[34,35]. For simplicity, we label these pairs of modes from hereon as $TO_2$ and $TO_3$, respectively. These higher frequency modes, in contrast to the soft mode, cannot be directly driven by the THz pulse because they lack spectral overlap. Instead, we propose that they are driven indirectly by coupling to the soft mode which is driven so far from equilibrium as to produce new frequency components because of its large anharmonicity. Further, the $TO_3$ mode in bulk STO is a *silent* mode that has neither infrared nor Raman cross section and can only be observed through higher order spectroscopy methods such as hyper-Raman[36].

The dynamics of the excitation process can be illustrated by calculating the time-resolved change in spectral content. In Figure 2c we present a time-frequency plot of the x-ray data in Fig. 2a using a 2.5 ps full-width, half-maximum sliding window fast Fourier transform (FFT). The red shaded area is a contour (30% of peak magnitude) of the same analysis for the $E_{THz}$ trace. While there is clear overlap at early times between the x-ray and THz responses, there are persistent frequency components in three bands: 1-2 THz (soft mode band), 4.5-5.5 THz ($TO_2$ band), and 7-8 THz ($TO_3$ band). Within the resolution of the time window, both high frequency bands begin to oscillate at the same time, immediately after the main low frequency response.

To test our hypothesis of nonlinear coupling through the soft mode phonon, we tuned the soft mode phonon frequency in and out of resonance with the driving THz field by varying the sample temperature. The THz peak field strength was kept constant at 880 ± 50 kV/cm. Figure 3a shows the sample optical conductivity (measured by THz ellipsometry[27]) as a function of temperature overlaid with the spectrum of $E_{THz}$. The mode frequency changes from 1.5 THz at 100 K to >2.5 THz at 250 K and thus moves out of resonance with the spectral components of $E_{THz}$. In Fig. 3b we show the magnitude of the FFT of the trXRD signal as a function of temperature, normalized to the peak of the 100 K trace. As we move out of resonance with the THz pulse (increasing temperature), the spectral weight between 1-2 THz diminishes, consistent with our model of resonant excitation. Furthermore, the peaks at 5.15 THz and 7.6 THz disappear as the temperature is increased, because the soft mode amplitude is no longer large enough to couple to these modes.

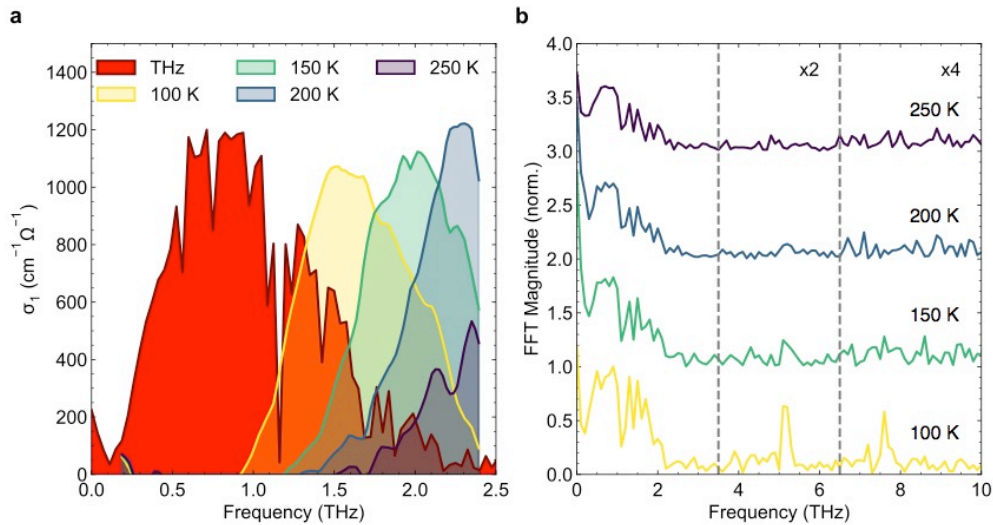

Figure 3: Temperature dependence of sample response. a, Optical conductivity data measured by THz ellipsometry show the temperature dependence of the soft phonon mode. Only at the lowest temperatures is there significant overlap with the THz pump spectrum (red area, normalized units) to drive the soft mode into the nonlinear regime. b, Fourier transform of trXRD signal showing transfer to high frequency modes at the corresponding temperatures in a.

Alternatively, we can tune the energy coupled to the soft mode by adjusting the THz field strength. Here we studied the sample response while the incident THz pulse was continuously attenuated with a pair of wiregrid polarizers without affecting the pulse shape or spectrum. The highest field value was achieved by removing the polarizers altogether. The trXRD signal with increasing THz field strength is shown in Fig. 4a. Several hallmarks of nonlinear phenomena are present: the trXRD peak signal saturates as a function of THz field strength, manifest as a sublinear relation between peak signal and applied field (Fig. 4b); the rise of the primary peak (near 1.8 ps) also grows steeper with increasing THz field, indicating new frequency content generated through a nonlinear structural response; at the very highest fields we observe the onset of high frequency oscillations superimposed upon the primary signal. For clarity, in Fig. 4c we emphasize the high frequency response by passing a subset of the data in Fig. 4a through a 3.5 THz high-pass filter. We see a turn-on of the high frequency oscillations at approximately 1.5 ps, near the first minimum of the trXRD response, suggestive that the large amplitude motion of the soft mode enables energy transfer into the higher frequency components. In Fig. 4d we present the trXRD spectrum for the same peak THz fields as in Fig. 4c, indicating the scaling of the $TO_2$ zone center mode at 5.15 THz.

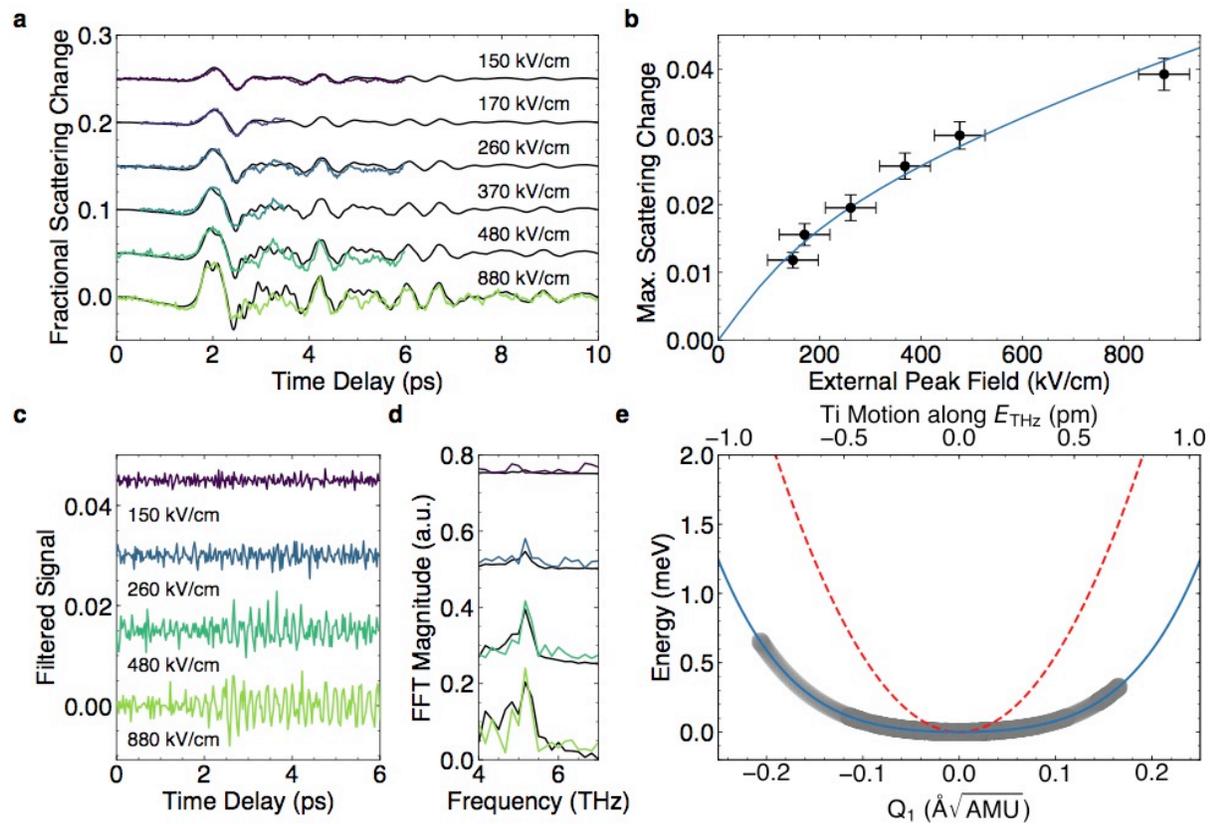

Figure 4: THz field dependent data and model calculations. a, Time-resolved change in x-ray scattering intensity for several THz pulse peak electric field values. Model data are shown as black solid lines. b, Maximum x-ray intensity change at increasing THz peak field strength (black dots) illustrating the saturation effect due to the quartic nature of the potential. The solid line is our model. c,d, High frequency portion (>3.5THz) of our data illustrating the onset of the $TO_2$ phonon oscillations in the time domain (c) and frequency domain (d). e, Phonon potential for soft mode used for simulations in a (blue) overlaid with harmonic potential at the same soft mode frequency (red, dashed). The gray shading represents the trajectory of the simulated soft mode at largest applied field (880 ± 50 kV/cm).

The experimental features of our data can be understood with a model based on a driven anharmonic oscillator with coupling to other, higher frequency oscillators. For simplicity, we consider only the soft mode phonon and the TO₂ phonon and neglect the silent mode. We label the soft mode and TO₂ modes respectively as $Q_1$ and $Q_2$, and thus the lattice potential can be expressed as follows:

$$V = \frac{1}{2}\omega_1^2 Q_1^2 + \frac{1}{2}\omega_2^2 Q_2^2 + \frac{1}{4}\kappa_1 Q_1^4 + \frac{1}{4}\kappa_2 Q_2^4 + \frac{1}{2}\chi Q_1^2 Q_2^2 + \psi_{12} Q_1^3 Q_2 + \psi_{21} Q_1 Q_2^3. \quad (1)$$

Here, $\omega_i$ denote the eigenfrequencies and $\kappa_i$, $\psi_{ij}$ and $\chi$ are the anharmonic phonon and phonon-phonon contributions to the potential. By symmetry only even terms are included in the potential. The time dependent dynamics are obtained from equation (1) by considering the equations of motion as:

$$\ddot{Q}_i + \frac{dV}{dQ_i} + \gamma_i \dot{Q}_i = Z_i^* \tilde{E}_{\text{THz}}. \quad (2)$$

We include additional terms containing $\gamma_i$ to account for the lifetime of phonons and $Z_i^* \tilde{E}_{\text{THz}}$ to incorporate coupling to the driving field $\tilde{E}_{\text{THz}}$ via the mode effective charge $Z_i^*$ [37] and adjusted for the effective field in the sample through $\tilde{E}_{\text{THz}} = \beta E_{\text{THz}}$, where $\beta$ quantifes the screening of the vacuum THz field by the sample. The solution of these coupled differential equations yields the dynamics of each individual mode for a given field, and the total structural dynamics is given by the superposition of the two eigenvectors.

To explore the specific dynamics for our STO film, we utilize first principle calculations to quantify the sizes of the anharmonic coupling terms and the mode effective charges, employing a previously established approach based on density functional theory[38,39]. We used ellipsometry measurements of our sample (Fig. 3a) to obtain the frequency and lifetime of the soft mode and used literature values for the $TO_2$ mode [32,33]. Because the calculations are performed at 0 K, we tuned several of the more sensitive parameters from equations (1,2) ($\kappa_1, \psi_{12}, \chi, \beta$) to best match our experimental results at 100 K. We extract the unit cell structure and eigenvectors from our first principle calculations and use these results to transform the calculated phonon amplitudes into the expected x-ray scattering change of the (2 -2 3) reflection. Numerical details are given in the Methods section and supplement.

The results of our model in comparison with the experimental data are shown in Fig. 4. We readily reproduce the primary time-domain features (black lines in Fig. 4a), including the overall waveform shape as well as the peak saturation (Fig. 4b) and appearance of 5.15 THz mode (Fig. 4d). Figure 4e shows the resultant quartic potential for the soft mode $Q_1$ (blue, solid) overlaid with the quadratic potential present when all nonlinear terms are set to zero (red, dashed) for comparison. The gray shaded region indicates the maximum excursion in our simulation when using the highest experimental THz field strength (880 ± 50 kV/cm), reaching about $Q_1 = -0.2 \text{ Å}\sqrt{AMU}$. This corresponds to 1.3 pm real space motion parallel to the THz driving field for the Ti ions (see supplement for full atomic displacement information). For reference, in the prototypical ferroelectric perovskite $BaTiO_3$, the Ti ion moves 9 pm from the cubic site at 300 K[40].

We can gain more physical insight into the dynamics of the system by further simplifying our model. The oscillator strength for the soft mode is much larger than the $TO_2$ mode, hence $Q_1 \gg Q_2$ while the THz driving field is present. Moreover, energy transfer into this mode is favored because $Q_1$ is in resonance with the THz field. Therefore, the most dominant nonlinear driving terms in the equations of motion must be $-\kappa_1 Q_1^3$ and $-\psi_{12} Q_1^3$ for $Q_1$ and $Q_2$, respectively. This allows us to simplify the equations of motion (2) as

$$\ddot{Q}_1 + \gamma_1 \dot{Q}_1 + \tilde{\omega}_1^2 Q_1 = Z_1^* \tilde{E}_{THz}, \qquad (3a)$$

$$\ddot{Q}_2 + \gamma_2 \dot{Q}_2 + \omega_2^2 Q_2 = Z_2^* \tilde{E}_{THz} - \psi_{12} Q_1^3, \qquad (3b)$$

where $\tilde{\omega}^2 = (\omega_1^2 + \kappa_1 Q_1^2)$ describes an amplitude-dependent frequency. Equation (3a) thus describes the motion of a driven oscillator with a frequency that increases with mode amplitude. This is consistent with the soft mode hardening effect observed in STO in static electric fields[34]. We attribute the saturation and steepening of the peak trXRD signal to this effect: for large amplitudes, the resonance condition for $Q_1$ shifts above the bandwidth of the THz driving field, thereby reducing energy transfer from the THz field into the soft mode as $Q_1$ increases.

For the equation of motion of $Q_2$, equation (3b), we neglect any nonlinear terms. Equation (3b) then corresponds to a harmonic oscillator with two driving terms: the THz electric field $\tilde{E}_{THz}$ and the other phonon mode $Q_1$. The latter term contains the third power of $Q_1$ which results in the generation of additional spectral components at the third harmonic of the spectrum of $Q_1$. This extends the fundamental bandwidth of the driving THz field to overlap with the $Q_2$ mode and results in more efficient energy transfer. This simple picture is in excellent agreement with our observation of the 5.15 THz TO$_2$ resonance driven by the anharmonic soft mode. We find that we best reproduce our data with $\kappa_1$ = 8750 THz$^2$Å$^{-2}$AMU$^{-1}$, $\psi_{12}$ = -3850 THz$^2$Å$^{-2}$AMU$^{-1}$, and $\chi$ = 3900 THz$^2$Å$^{-2}$AMU$^{-1}$. For reference, Katayama et al. use a single quartic potential to model the soft mode in STO films, yielding the equivalent of $\kappa$ = 19000 THz$^2$-Å$^{-2}$-AMU$^{-1}$ in reasonable agreement with our results[41]. With only two nonlinear terms therefore we can capture the field-dependent observations in our x-ray scattering signal.

It is striking that we observe a response at the 7.6 THz mode because in bulk STO this phonon is silent (neither IR nor Raman active)[36]. Ellipsometry measurements on our thin-film sample at room temperature show no IR activity in this frequency window (see supplement). The anharmonicity described by our model provides a ready channel to couple energy from one phonon mode to another and could be expanded to incorporate this additional mode. Thus, our novel excitation method has the potential to stimulate otherwise inaccessible phonon modes via optical techniques.

In conclusion, we have reported direct structural evidence of phonon up-conversion in the perovskite incipient ferroelectric STO. Intense single-cycle THz radiation couples directly to the soft mode via infrared absorption and drives the system far from equilibrium into the strongly nonlinear regime. Subsequent energy coupling into high frequency phonon modes at 5.15 and 7.6 THz occurs, as directly observed in time-resolved x-ray diffraction measurements. We can capture the nonlinear response by invoking terms up to fourth-order coupling between the soft mode and the 5.15 THz mode, reproducing with high fidelity the observed x-ray scattering signal.

This novel excitation mechanism provides an avenue to transfer energy into higher energy modes because it exploits the nonlinearities that couple phonons. It can even be used to channel energy into so-called "silent modes" that are neither IR- nor Raman-active. Moreover, current tools for enhancing and engineering THz interactions with matter[28,42,43] may now be leveraged to control higher-frequency, mid-infrared modes via this nonlinear phonon up-conversion process.

**Acknowledgements**

Use of the Linac Coherent Light Source (LCLS), SLAC National Accelerator Laboratory, is supported by the U.S. Department of Energy, Office of Science, Office of Basic Energy Sciences, under Contract No. DE-AC02-76SF00515. M.K. and M.C.H. are supported by the U.S. Department of Energy, Office of Science, Office of Basic Energy Sciences, under Award No. 2015-SLAC-100238-Funding. U.S. acknowledges support from the National Center of Competence in Research: Ultrafast Science and Technology (NCCR MUST) of the Swiss National Science Foundation. S.B. acknowledges support from the Knut and Alice Wallenberg Foundation. M.K. and M.C.H. extend thanks to W. Chueh and A. Baclig for annealing the sample. M.F. extends thanks to M. Först for fruitful discussions about modelling the STO system.

**Author Contributions**

M.K. and M.C.H. conceived the experiment and performed the final data analysis. M.F. provided DFT calculations and theory support. T.v.D. and S.B. helped with on-line data analysis. M.K., M.C.H., J.M.G. and D.Z. performed the time-resolved x-ray experiment. U.S. provided sample expertise and additional x-ray data. M.R. prepared the sample. P.M. and C.B. made the THz ellipsometry measurements of the sample. The paper was written by M.K. and M.C.H., with discussions from other authors.

**Competing Interests**

The authors declare no competing interests.

## Methods

**Experimental Details.** The measurements were carried out at the XPP instrument of the Linac Coherent Light Source[44] in monochromatic mode. The x-rays were tuned to 8 keV with a bandwidth of ~1 eV after the monochromator[45] and were 30 fs full-width, half-max (FWHM) in duration at 120 Hz repetition rate.

Single-cycle THz radiation was generated through optical rectification of 800 nm femtosecond pulses in LiNbO$_3$ using the tilted pulse-front technique [46,47]. We used up to 20 mJ pulse energy at 120 Hz repetition rate and 100 fs FWHM pulse duration[48]. The THz pulses were focused into the sample using a three parabolic mirror geometry with intermediate focus. A pair of wiregrid polarizers was used to controllably attenuate the THz field strength without affecting the pulse shape by rotating the first polarizer. We characterized the electric field component of the THz pulse at the sample location using electro-optic sampling in 50 μm thick (110) cut GaP crystals and found the peak electric field to be 880 ± 50 kV/cm. The peak frequency of the THz pump pulse is at 0.75 THz with components extending to 2.5 THz.

The intrinsic timing jitter between the THz pump and x-ray probe pulses was mitigated using a spectral encoding technique[49], and data were temporally binned to 25 fs resolution. We collected x-ray scattering data with the CSPAD area detector[50,51] and normalized with a shot-by-shot x-ray intensity monitor as well as by shots when the pump laser was not present. A typical dataset included 60k (36k) x-ray shots for the laser on (off) after excluding shots with low x-ray intensity or poor spectral encoding signal. We fixed the detector and sample in one scattering geometry and integrated over a region of interest on the detector to estimate the diffraction peak intensity. Error bars on the trXRD signal were estimated from the standard deviation of the signal for each x-ray shot.

We maintained the sample at ~100 K using a nitrogen cryostream (Oxford Instruments Cryojet 5). The sample temperature thus has a lower bound of 100 K but could be at most 10 K higher because the temperature is measured at the cyrostream jet output.

**Sample Details.** Our sample consisted of a 50 nm thick STO film on an $(LaAlO_3)_{0.3}(Sr_2TaAlO_6)_{0.7}$ (LSAT) substrate. It was annealed at 1200 C for 12 h to diminish the effects of the substrate on the phonon dispersion[27,52]. X-ray diffraction measurements before annealing indicated the film was compressed from bulk STO in-plane to match the substrate and expanded in the cross-plane direction. The lattice parameters were $a_0$=3.867Å and $c_0$=3.925Å. See supplement for static x-ray diffraction data.

**Structure Factor Calculations.** Experimentally we measure the diffraction peak intensity $I$ for a given set of lattice planes corresponding to the reciprocal lattice vector $\boldsymbol{G}$. As the ions in the lattice respond to the THz excitation, the x-ray scattering intensity changes. The intensity $I \propto |F|^2$, where the structure factor $F = \sum_j f_j e^{-i\boldsymbol{r}_j \cdot \boldsymbol{G}}$. Here $f_j$ and $\boldsymbol{r}_j$ are the atomic scattering factor and position vector for the ion $j$, respectively; the sum runs over all ions in the unit cell. Note that the $f_j$ are complex and extrapolated from tabulated values[53–55] for the experimental x-ray energy of 8 keV. In our model we calculate the phonon amplitudes $Q_i$, which we use to extract the change in ion position through the mode eigenvectors $\xi_{ijk}$ via $r_{jk} = r_{jk_0} + \sum_i \xi_{ijk} Q_i$, where $i, j, k$ index respectively the phonon branch, ion, and coordinate direction, and $r_{jk_0}$ is the equilibrium position of the ion. Denoting $F_0$ as the structure factor in equilibrium, the fractional change in x-ray scattering intensity $\Delta I/I = |F|^2/|F_0|^2 - 1$. We calculate the STO unit cell structure $r_{jk_0}$ and eigenvectors $\xi_{ijk}$ using density functional theory as described below in the Numerical Details and combine this information with the phonon amplitudes $Q_i$ to compute the expected relative change in x-ray scattering intensity.

**Numerical Details.** Our calculations were carried out using density functional theory (DFT) as implemented in the Quantum Espresso code[56]. We used projected augmented wave (PAW) pseudopotentials, which contain as valence states the $4s^2 5s^2 4p^6$ for Sr, $3s^2 3p^6 3d^2 4s^2$ for Titanium and $2s^2 2p^4$ for Oxygen. In all computations, we sampled the Brillouin zone by a 14x14x14 k-point mesh generated with the Monkhorst and Pack scheme[57] and placed a cutoff energy on the wavefunction of 45 Rydbergs. As an approximation for the exchange correlation functional, we applied PBEsol[58]. All total energy computations have been reiterated until the change in energy became less than $10^{-10}$ Rydberg. As phonon computations require a force-free groundstate, we performed structural relaxation until the forces acting on individual ions became lower than 5 μRy/$a_0$. The phonon frequencies and eigenvectors have been obtained by density functional perturbation theory[59] calculations, whereas the mode effective charges are computed utilizing the modern theory of polarization[37,60]. Finally, we compute the anharmonic coupling coefficients from frozen phonon calculations. Thereby, we modulate the structure with an appropriate superposition of the phonon eigenvectors and compute the resulting total energies. The anharmonic coefficients listed in equation (1) are then obtained by a least mean squares fit of the multidimensional potential landscape. We note that in our fitting procedure we also take into account higher order terms than those given in equation (1), which due to their size, however, for the specific case of STO can be neglected. In order to fit our experimental data, we tweaked several of the anharmonic coefficients as noted in the main text, recognizing that the DFT calculations are performed at 0 K while our experimental measurements were collected ~100K. Numerical values for the STO unit cell structure, eigenvectors, and model coefficients from equations (1-3) are listed in the supplement.

# Supplementary material

Kozina et al, "Terahertz-Driven Phonon Upconversion in $SrTiO_3$"

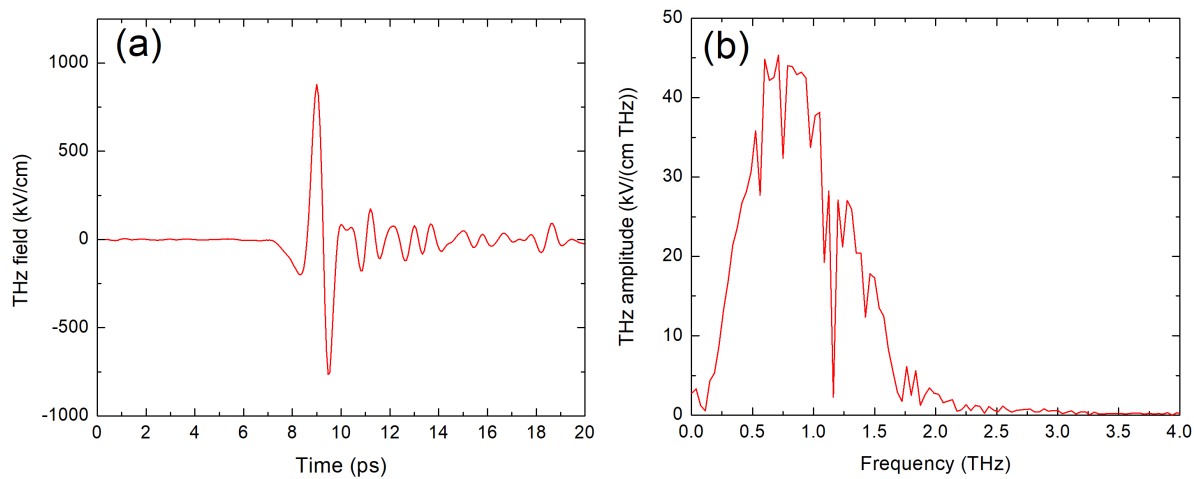

**Figure S1 Terahertz pump waveform:** *a Electro-optic sampling signal of terahertz (THz) pump at sample location measured in-situ by electro-optic sampling in a 50 μm thick (110) GaP crystal. b Amplitude spectrum of the THz waveform in a. Intensity adjusted with wiregrid polarizers (Infraspecs P01) without changing THz waveform.*

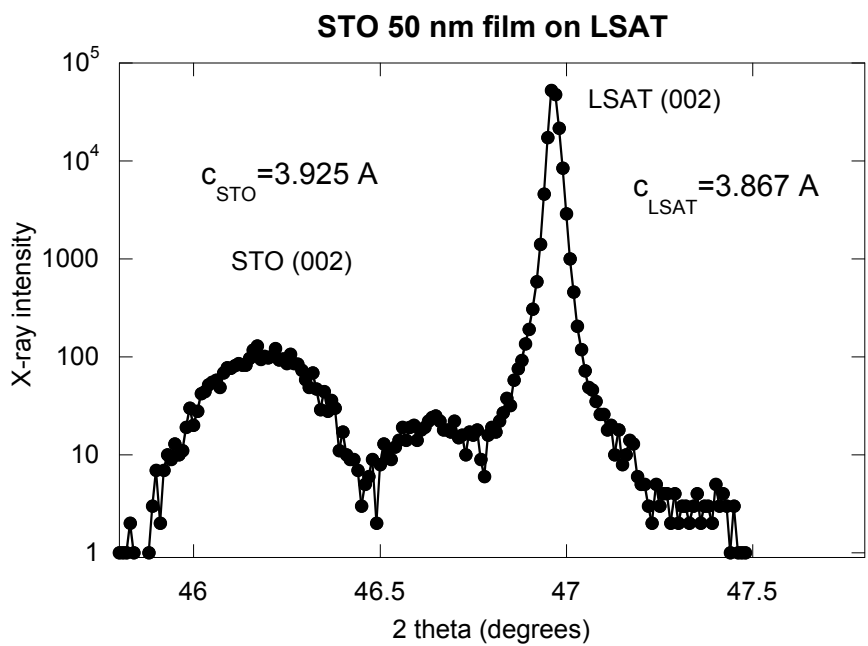

**Figure S2: Two-Theta scan of sample before annealing.** *Static x-ray diffraction pattern along sample normal. The low-angle peak corresponds to the STO film and the sharp peak at higher angle is the LSAT substrate. We observe clear splitting along the cross-plane direction between the substrate and the film. These measurements were taken before the sample was annealed.*

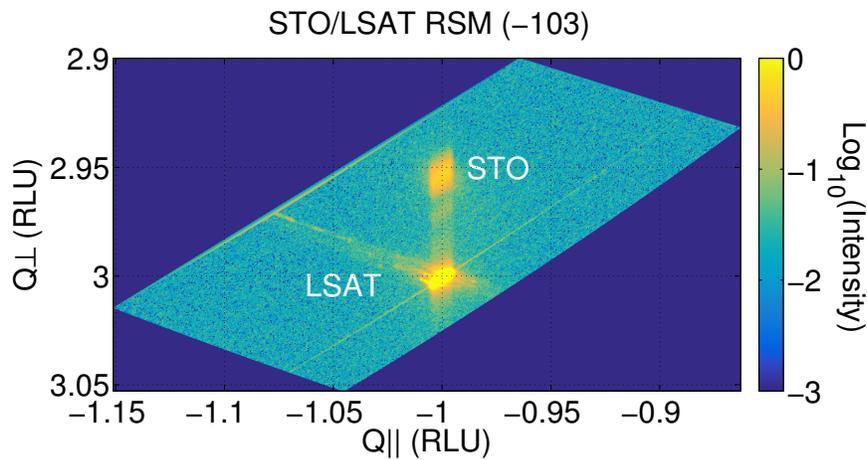

*Figure S3: Reciprocal space map of the sample before annealing.* Static reciprocal space map collected on a similar sample prior to annealing about the (-1 0 3) reciprocal lattice point using a Cu kα source. The separation between the film and substrate peaks along the cross-plane direction shows there is a difference in lattice parameter, whereas there is clear indication that in-plane the film and substrate are lattice matched to the LSAT unit cell.

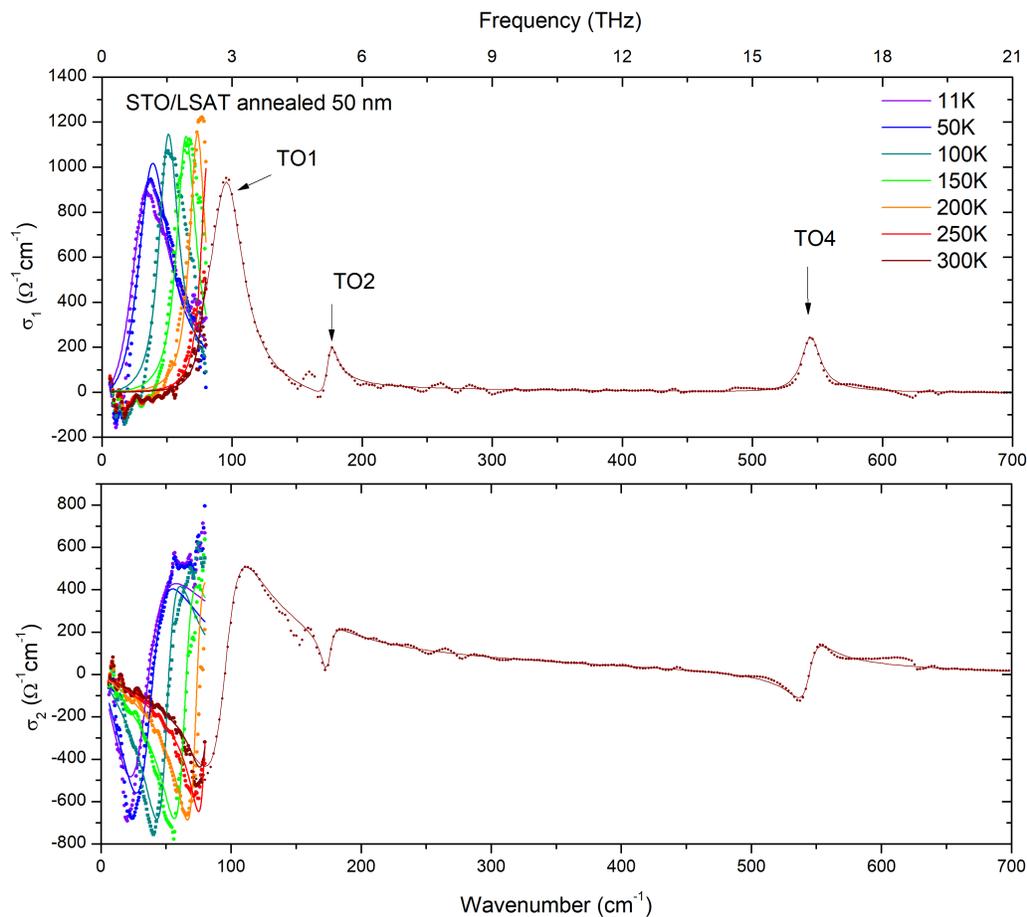

*Figure S4: Optical conductivity of the annealed sample.* We show additional ellipsometry measurements on our sample incorporating extra temperature points at low and high temperature. The extended frequency range at 300K shows the $TO_2$ and $TO_4$ phonon modes. Data from Fig. 3a are also included for comparison.

|   | Sr | Ti | Oxy | Ozx | Oyz |
|---|---|---|---|---|---|
| X | 0 | 0.5 | 0.5 | 0.5 | 0 |
| Y | 0 | 0.5 | 0.5 | 0 | 0.5 |
| Z | 0.00987045 | 0.51190984 | 0.9970011 | 0.4956093 | 0.4956093 |

**Table S1 Unit Cell Relative Coordinates from DFT.** *Lattice is tetragonal with $a_0$ = 3.89000 Å and $c_0$ = 3.91043 Å.*

|   | Sr | Ti | Oxy | Ozx | Oyz |
|---|---|---|---|---|---|
| X | 0.02963898 | 0.04544241 | -0.08189972 | -0.09030182 | -0.07348941 |
| Y | -0.02957567 | -0.04530107 | 0.08204853 | 0.07359076 | 0.09041028 |
| Z | 0 | 0 | 0 | 0 | 0 |

**Table S2 Eigenvector for $Q_1$ mode.** *Eigenvector calculated with $E_{THz}$ parallel to (X, -Y, 0). The units of the eigenvector are Å*AMU$^{1/2}$. The eigenvector is normalized so that $\sum \xi_{ijk}^2 m_{jk} = 1$. Here $\xi_{ijk}$ and $m_{jk}$ are the eigenvector and ion mass in AMU, and the indices i,j,k label the eigenvector mode, ion, and coordinate direction.*

|   | Sr | Ti | Oxy | Ozx | Oyz |
|---|---|---|---|---|---|
| X | -0.04831474 | 0.07770832 | 0.01575631 | 0.00874036 | 0.00874793 |
| Y | 0.04829799 | -0.07773579 | -0.01568512 | -0.00870886 | -0.00868356 |
| Z | 0 | 0 | 0 | 0 | 0 |

**Table S3 Eigenvector for $Q_2$ mode.** *Eigenvector calculated with $E_{THz}$ parallel to (X, -Y, 0). The units of the eigenvector are Å*AMU$^{1/2}$. The eigenvector is normalized so that $\sum \xi_{ijk}^2 m_{jk} = 1$. Here $\xi_{ijk}$ and $m_{jk}$ are the eigenvector and ion mass in AMU, and the indices i,j,k label the eigenvector mode, ion, and coordinate direction.*

| | | | | | |
|---|---|---|---|---|---|
| $\omega_1/(2\pi)$ | 1.669 THz | $\kappa_1$ | 8750 THz$^2$Å$^{-2}$AMU$^{-1}$ | $\chi$ | 3900 THz$^2$Å$^{-2}$AMU$^{-1}$ |
| $\omega_2/(2\pi)$ | 5.156 THz | $\kappa_2$ | **1066 THz$^2$Å$^{-2}$AMU$^{-1}$** | $\psi_{12}$ | -3850 THz$^2$Å$^{-2}$AMU$^{-1}$ |
| $\gamma_1/(2\pi)$ | 0.900 THz | $Z_1^*$ | **2.6 e$^-$AMU$^{-1/2}$** | $\psi_{21}$ | **-3363 THz$^2$Å$^{-2}$AMU$^{-1}$** |
| $\gamma_2/(2\pi)$ | 0.150 THz | $Z_2^*$ | **0.2 e$^-$AMU$^{-1/2}$** | $\beta$ | 0.21 |

**Table S4 Model parameters.** *Values in bold are unmodified from DFT calculations at 0 K. Values in red were extracted from spectroscopy and literature. Parameters $\kappa_1$, $\psi_{12}$, and $\chi$ were initially established by DFT and then tuned to best fit the experimental data. The parameter $\beta$ was informed by calculations for the expected field screening inside the STO film on the LSAT substrate but was fine-tuned to best fit the experimental data.*